\documentclass[aps,twocolumn,10pt,nofootinbib]{revtex4}
\usepackage{graphicx,amssymb,amsmath,subfigure}

\newcommand{\braket}[1]{\left<#1\right>}
\newcommand{\para}[1]{\left(#1\right)}

\newcommand{\ld}[0]{\ \ \ \ \ \ }

\begin{document}
\title{Fractional quantum Hall effect at zero magnetic field}
\author{Abolhassan Vaezi$^{*}$}
\affiliation{Department of Physics, Sharif University of Technology\\
P.O.Box 11365-9161, Tehran, IRAN \\
And \\
Institute for Studies in Theoretical Physics and Mathematics (IPM)\\
P.O.Box 19395-5531, Tehran, IRAN
}
\email{vaezi@mit.edu}

\date{\today}

\begin{abstract}
In this letter, we discuss the recently proposed fractional quantum Hall effect in the absence of Landau levels. It is shown that the parton construction can explain all properties of 1/3 state, including the effective charge of quasiparticles, their statistics and the many-body ground-state degeneracy. The low energy description of these states has been discussed. We also generalize our model to construct the hierarchical quantum Hall states at filling fractions other than $1/m$.
\end{abstract}

\maketitle

\section{Introduction}

Recently, several authors have tried to mimic the fractional quantum Hall effect (FQHE) \cite{Laughlin_1983_1} in lattice
models, but in the absence of external magnetic field and Landau levels \cite{Chamon_2010_1,E_Tang_2010_1,ZCG_2011_1,ZCG_2011_2}. To do so, they
considered a gapped electron system with a nearly
flat lowest energy band. Nearly flat band means that the ratio of the bandwidth to the energy gap is small. Despite the conventional quantum Hall systems where
the magnetic field produces flat bands (Landau levels) with
nontrivial topology i.e. nonzero Chern number, and also determines the degeneracy of each band,
in this case the degeneracy is imposed by the lattice structure i.e. the
number of momenta in the Brillouin zone and the
nontrivial topology(non-zero Chern number) originates form the
time-reversal symmetry breaking through threading a nonzero flux in different
Wilson loops (by assuming complex hopping integrals). A famous example of this kind at $v=1$ filling fraction, was proposed by Haldane in 1988 \cite{Haldane_1988_a}. When the band is totally filled by
electrons ($\nu=1$), the bulk of the system is a band insulator, and it can be studied through the standard
non-interacting picture and everything including the Chern number is
well-defined because of the gapped nature of the system. But what
if only a fraction of the lower band is occupied by electrons? Using
exact diagonalization methods for small size systems, it has been reported
that at $\nu=1/3$ filling fraction, similar to the 1/3 Laughlin state, the
Chern number of the many-body state is 1/3, the ground-state degeneracy
is 3 and therefore the charge of quasiparticles is e/3 \cite{Chamon_2010_1}. We also expect quasiparticles to be anyons with $\theta=\frac{\pi}{3}$ statistics. These together implies that the low energy description of these systems is identical to that of the Laughlin $\nu=\frac{1}{3}$ state \cite{Wen_QHE_1992_1,Wen_QHE_1999_1,Levin_2009_1}.

Now, let us discuss the difference between the above mentioned state with the Haldane's model. Haldane assumed a
gapped state (i.e. the many body ground state is separated by a gap
from all excited states), so it starts from a state whose bulk is
insulating and the filling fraction is one. Since the Brillouin zone can be
viewed as a 2D torus, and we have a mapping from this torus to a 2D
sphere (because we have two bands and the Hamiltonian can be expanded in terms of Pauli matrices. The coefficients of these matrices represent a vector in the momentum space and the corresponding unit vectors live on a 2D sphere). Since all momenta on the torus are
occupied by quasi-particles, the homotopy group of this mapping
is well-defined and we can identify it with the first Chern number of the system under consideration.
As a result, the Hall conductance is nonzero in that case. But in our case, the
question is that can we have something similar to FQHE at other filling
fractions? For example, at 1/3 filling fraction, since only a fraction of the
lower band is occupied by electrons, if we ignore interaction effects, the non-interacting many body ground-state is
gapless and the band theory predicts a metallic behavior in this case.
On the other hands, the state is gapless and the Chern number is ill-defined. It is worth noting that because we are in commensurate filling fraction, the charge density wave may gap out the system and electrons will live on a gapped band, but in that case, because of the non-interacting nature of the resulting state, the Chern number of that state would be an integer \cite{Thouless_1982_1} and the ground state degeneracy will be one, while in numerical studies, the Chern number is 1/3 and the ground state degeneracy is 3. Therefore the charge density or any other noninteracting approach cannot solve the problem. It is the strong interacting nature of the system that leads to such exotic behaviors. 

After taking the effect of interaction between electrons into consideration, a
nonzero gap will open up between the many body ground-state and excited states, and therefore the Chern number becomes
well-defined. So far, everything was similar to the ordinary FQHE.
But what are differences? To answer this question, we need to look more carefully at the parton construction in the presence of
an external magnetic field. Let us assume that $N_{\phi}$ (number of
flux quanta) = 3 Ne (number of electrons), so we are at $\nu=1/3$ filling
fraction. It should be noted that it is the magnetic field which
determines the degeneracy of each Landau level(LL). This degeneracy for
charge $q$ quasi-particles is $g(q)=\frac{\Phi}  {\Phi_{0}\para{q}}=q\times N_{\phi}$ where $\Phi=BL_xL_y$ is the total
flux and $\Phi_{0}(q)=2\pi / q$ is the quantum of magnetic flux assuming
the charge of quasi-particles is $q$ . So the degeneracy $g(q)$ is
proportional to the charge $q$ . Let us assume each electron is formed
of three partons i.e. $C_{i}=f_{1,i}* f_{2,i}* f_{i,3}$, and each parton has charge $q=\frac{e}{3}$.
So the degeneracy of Landau Levels for each parton is $g(e/3)=\frac{g(e)}{3}=\frac{N_{\phi}}{3}=\frac{3 Ne}{3} = Ne$. Since
the number of each flavor of these partons equals the number of
electrons, we have  $g(e/3)=N_e=N_{f_1}=N_{f_2}=N_{f_3}$, so every flavor
is at $\nu=1$ (filled Landau level) and their Chern number is one. The
wavefunction of electrons is the product of the wavefunction of these
partons but the number of all flavors should be equal at every site.
So the wave-function of electrons is the cube of the wavefunction of
the filled Landau level of charge $e/3$ particles, i.e.
$\psi_e=\psi_{f_1}~^3$ = Laughlin state for $\nu=1/3$. So the key point is
that, {\bf it is the combination of the magnetic field and  the charge of quasiparticles that determine the degeneracy of Landau levels}. So
using the parton construction will lead to three copies of filled
Landau levels and in that way we obtain the Laughlin state. It is easy to show that the low energy description of this systems using parton construction is the following Chern Simons action which is identical to the low energy theory of the $\nu=1/3$ Laughlin wave function:

\begin{eqnarray}
  \mathcal{L}=\frac{3}{4\pi} \epsilon^{\mu\nu\rho} a_{\mu} \partial_{\nu} a_{\rho} + \frac{1}{2\pi}\epsilon^{\mu\nu\rho} A^{EM}_{\mu} \partial_{\nu} a_{\rho}
\end{eqnarray}

In our case however, because of lattice effects we have a highly degenerate lowest band (i.e. we have an almost flat band). Therefore it is {\bf NOT} the magnetic field or the charge of
quasiparticles that defines the ground-state degeneracy. {\bf The degeneracy of the lowest non-interacting state
is actually determined  by lattice structure instead of magnetic flux.}
Therefore, in the absence of external magnetic
field, lattice effects are much more important.

\section{Parton construction for $\nu=\frac{1}{3}$ filling fraction}
Let us consider $v=\frac{1}{3}$ case, i.e. the number of electrons is 1/3 of the number of $k$ points in the Brillouin zone. In this paper we only consider spinless electrons and will ignore their spin degree of freedom. We assume that the bandwidth of the lowest band is much smaller than the energy gap, so we can consider the lowest band as a highly degenerate state. We also assume that electrons are interacting with each other with a coupling constant larger than the band width but smaller than the energy gap, so the system becomes strongly interacting. Within parton construction method, the electron annihilation operator is written in terms of fermionic parton operators as follows:

\begin{eqnarray}\label{Eq_C_def}
  C_{i}=f_{1,i}f_{2,i}f_{3,i}=\frac{1}{6}\epsilon^{\alpha,\beta,\gamma}f_{\alpha,i}f_{beta,i}f_{\gamma,i}
\end{eqnarray}

along with two local constraints on the number of partons which are

\begin{eqnarray}\label{Eq_cons_1}
&&f_{1,i}^\dag f_{1,i}=f_{2,i}^\dag f_{2,i}\cr
&&f_{2,i}^\dag f_{2,i}=f_{3,i}^\dag f_{3,i}
\end{eqnarray}

These constraints projects the Hilbert space of partons to the physical Hilbert space of electrons. The parton construction method is actually a special kind of slave particle technique \cite{Barkeshli_2010_1,Barkeshli_2010_2,Vaezi_2010a,Vaezi_2011_a}. Let us assume the charge of these partons to be $q=\frac{e}{3}$. It is obvious from Eq. \ref{Eq_C_def} that $C_{i}$ operator is invariant under local $SU(2)$ gauge transformations. The parton construction becomes justified in the presence of strong interactions. Now let us start from the following hopping Hamiltonian for physical electrons:

\begin{eqnarray}
  H_{0}=-\sum_{i,j}t_{i,j}C_{i}^\dag C_{j}
\end{eqnarray}

When the lowest band is fractionally filled by electrons, interaction effects become more important and we need to add an interaction term of the form $\sum V_{i,j} n_{i} n_{j}$ to the above Hamiltonian to include Coulomb repulsion between electrons. Now, let us define the Berry connection in the following way:

\begin{eqnarray}
  A_{\vec{k}}=\sum_{l \in \mbox{occupied state}}\left<k,l\right|i\partial_{\vec{k}}\left|k,l\right>
\end{eqnarray}

where $l$ labels different energy bands. Berry connection implies Berry curvature as follows:

\begin{eqnarray}
  F_{kx,ky}=\partial_{kx}A_{k_y}-\partial_{ky}A_{k_x}
\end{eqnarray}

We choose $t_{i,j}$'s such that the lowest band becomes as flat as possible and the integral of Berry connection over all states on the lowest band becomes nontrivial and equal to $2\pi$. Therefore we have

\begin{eqnarray}
&&  C=\frac{1}{2\pi}\int_{BZ}F_{kx,ky} d^2k=1
\end{eqnarray}

where $C$ is the first chern number of the lowest energy band. Since $C=1$, the lowest band has nontrivial topology. In terms of parton operators we obtain the following Hamiltonian

\begin{eqnarray}
  &&H=-\sum_{i,j}t_{i,j}f_{3,i}^\dag f_{2,i}^\dag f_{1,i}^\dag f_{1,i}f_{2,i}f_{3,i}\cr
  &&+\lambda_{1}\para{i}\para{f_{1,i}^\dag f_{1,i}-f_{2,i}^\dag f_{2,i}}+\lambda_{2}\para{i}\para{f_{2,i}^\dag f_{2,i}-f_{3,i}^\dag f_{3,i}}\ld
\end{eqnarray}

where we have used $\lambda_{1,2}$ fields as Lagrange multipliers to implement constraints in Eq. \ref{Eq_cons_1}. No use the saddle point approximation to make the Hamiltonian simple and noninteracting. In the most symmetric case, we assume $\braket{f_{\alpha,i}^\dag f_{\beta,i}}=\delta_{\alpha,\beta}\tilde{t}_{i,j}$. In this case $\braket{\lambda_{1}}=\braket{\lambda_2}=0$. Therefore we have the following effective Hamiltonians:

\begin{eqnarray}
&&  H_{\alpha}=-\sum_{i,j}\tilde{t}_{i,j} f_{\alpha,i}^\dag f_{\alpha,i}
\end{eqnarray}

We have the same spectrum for all partons. Let us call the energy dispersion of the lowest band $\epsilon_{k}$ and energy eigenstate for each parton $F_{\alpha,k}^\dag$. The Berry phase for electrons can be shown that is related to that of partons as follows:

\begin{eqnarray}
  \theta_{B,e}=\frac{\theta_{B,f_1}+\theta_{B,f_2}+\theta_{B,f_3}}{3}
\end{eqnarray}

In the symmetric case, we have $\theta_{B,e}=\theta_{B,f_1}=\theta_{B,f_3}=2\pi$. Therefore the Chern number of the band structure of partons is equal to that of the band structure of electrons, so we have $C_1=C_2=C_3=C=1$. But the number of each parton equals the number of electrons, therefore $N_{f_1}=N_{f_2}=N_{f_3}=N_e$. So all partons are at $\nu=\frac{1}{3}$ filling fraction. Since all three partons have the same energy spectrum, each energy is three time degenerate. This degeneracy can be easily left by perturbations. For example the following perturbation can do so:
\begin{widetext}
\begin{eqnarray}
  H=\sum_{k}\left[\begin{array}{ccc}
              F_{1,K}^\dag & F_{2,K}^\dag & F_{3,K}^\dag
            \end{array}\right]\left[
                         \begin{array}{ccc}
                           \epsilon_{1,k} & \Delta_{k}& \Delta_{k}\\
                           \Delta_{k} & \epsilon_{k} &\Delta_{k}\\
                           \Delta_{k}&\Delta_{k} &\epsilon_{k} \\
                         \end{array}
                       \right]
  \left[\begin{array}{c}
              F_{1,K} \\
              F_{3,K} \\
              F_{3,K}
            \end{array}\right]
\end{eqnarray}
\end{widetext}

The energy of the lowest band is $E_{k}=\epsilon_{k}-2\Delta_{k}$, and the energy eigenstates are created by $\bar{f}_{k}^\dag=\frac{F_{1,k}^\dag+F_{2,k}^\dag+F_{3,k}^\dag}{\sqrt{3}}$. Assuming max$(E_{k})~<~$min$(\epsilon_{k})$, the lowest band is a gapped state. Moreover the number of quasiparticles on the lowest band is $N_{f_1}+N_{f_2}+N_{f_3}=3N_e=$number of sites. So the lowest band is at $\nu=1$ filling fraction. Therefore we can define the the Chern number for it. It is easy to see that the Berry connection for $\bar{f}_{k}$ band is related to that of partons and we have:

\begin{eqnarray}
  \bar{A}_{\vec{k}}=\frac{A_{1,\vec{k}}+A_{2,\vec{k}}+A_{3,\vec{k}}}{3}=A_{1,\vec{k}}
\end{eqnarray}

So we have

\begin{eqnarray}
  \bar{C}=\frac{C_{1}+C_{2}+C_3}{3}=1
\end{eqnarray}

Therefore we have one edge state for this completely filled gapped state. Introducing a U(1) $\bar{a}$ gauge field to represent the current of $\bar{f}$ quasiparticles, we have the following relation:

\begin{eqnarray}
  \bar{j}_{\mu}=\frac{\bar{C}}{2\pi}\epsilon^{\mu,\nu,\rho}\partial_{\nu}\bar{a}_{\rho}=\frac{1}{2\pi}\epsilon^{\mu,\nu,\rho}\partial_{\nu}\bar{A}_{\rho}
\end{eqnarray}

On the other hand, as we already showed, the density and the current of $\bar{f}$ quasiparticles is three times bigger than that of electrons, i.e. we have:

\begin{eqnarray}
  \bar{j}_{\mu}=3j_{e,\mu}=\frac{3}{2\pi}\epsilon^{\mu,\nu,\rho}\partial_{\nu}a_{\rho}
\end{eqnarray}
where the U(1) $a_{\mu}$ gauge field represent the current of physical electrons. Therefore we have:

\begin{eqnarray}
  \frac{3}{2\pi}\epsilon^{\mu,\nu,\rho}\partial_{\nu}a_{\rho}=\frac{1}{2\pi}\epsilon^{\mu,\nu,\rho}\partial_{\nu}\bar{A}_{\rho}
\end{eqnarray}

The above equation can be obtained from the following Lagrangian which is:

\begin{eqnarray}\label{Eq_L1}
  &&\mathcal{L}=\frac{3}{4\pi}\epsilon^{\mu\nu\rho}a_{\mu}\partial_{\nu}a_{\nu}+\frac{1}{2\pi}\epsilon^{\mu\nu\rho}\bar{A}_{\mu}\partial_{\nu}a_{\nu}
\end{eqnarray}

which is the same as the Chern Simons action that describes $\nu=1/3$ Laughlin state, except that instead of electromagnetic gauge field $A^{EM}_{\mu}$, we have used Berry connection of the $\bar{f}$ quasiparticles $\bar{A}_{\mu}$. So, electrons are at $\nu=1/3$ filling fraction, quasiparticles are anyons with $\theta=\frac{\pi}{3}$ statistics, anyons carry $q=\frac{e}{3}$ charge and the ground-state degeneracy is 3. This parton construction method can be generalized to other filling $\nu=1/m$ filling fractions by using the following construction:

\begin{eqnarray}
  C_{i}=\prod_{\alpha=1}^{m}f_{\alpha,i}
\end{eqnarray}

For this construction we would obtain the following Chern Simons level $m$ action:

\begin{eqnarray}\label{Eq_L1}
  &&S_{CS}=\int d^2x dt \frac{m}{4\pi}\epsilon^{\mu\nu\rho}a_{\mu}\partial_{\nu}a_{\nu}+\frac{1}{2\pi}\epsilon^{\mu\nu\rho}\bar{A}_{\mu}\partial_{\nu}a_{\nu} \ld
\end{eqnarray}

\section{Hierarchical FQH states at zero B field}

We can use the hierarchy method to construct other filling fractions. Using the standard notation, we obtain the following action:

\begin{eqnarray}\label{Eq_L1}
  &&S_{CS}=\int d^2x dt \frac{K_{ij}}{4\pi}\epsilon^{\mu\nu\rho}a_{i,\mu}\partial_{\nu}a_{j,\nu}+\frac{Q_i}{2\pi}\epsilon^{\mu\nu\rho}\bar{A}_{\mu}\partial_{\nu}a_{i,\nu}\ld
\end{eqnarray}

where $K_{i,j}$ is the $K$ matrix in K-theory and $Q$ the charge of each level \cite{Wen_2004_book}. Just like the usual FQHE, the charge of anyons is $Q^{\top}.K^{-1}.Q$, and the degeneracy of the groundstate is $\det{K}$. A generic quasiparticles in the above hierarchical fractional quantum Hall state, carries $l_{i}$ unit of the $i$-th level.  Therefore it couples to $a_{i}$ gauge fields in the following way:

\begin{eqnarray}
  l_{i}a_{i,\mu}j^{\mu}
\end{eqnarray}

Adding this term to the Lagrangian of the Chern Simons action and integrating out $a_{i,\mu}$ gauge fields, we can compute the statistics of these quasiparticles and will obtain:

\begin{eqnarray}
  \theta=\pi l^{\top} K^{-1} l
\end{eqnarray}
and similarly the effective electric charge of those quasiparticles will be:

\begin{eqnarray}
  q_{Q}=e Q^{\top} K^{-1} l
\end{eqnarray}

\section{Discussion and conclusion}

The main difference between the usual FQHE and the FQHE at zero external magnetic field and driven by lattice effects, is that in the presence of external magnetic field, we obtain almost flat Landau levels, and the degeneracy of these levels is given by the total magnetic flux and the charge of quasiparticles. At $\nu=1/m$ filling fractions, where only $1/m$ of the lowest Landau level is occupied by electrons, we can assume electrons are composite objects that are formed up of $m$ smaller partons each of which carries $1/m$ charge of electrons. Assuming so, it is easy to check that each parton state is at filled Landau level and therefore the problem can be tackled easily. In the absence of external magnetic field, the degeneracy of the lowest band is given by the lattice structure (area of the BZ). Therefore assuming smaller quasiparticles, does not immediately solve the problem. We instead need to mix the sate of these $m$ partons to obtain a filled band with nonzero gap. In that case the Chern number of the ground state is well-defined. So we can write down an effective action for partons. Partons will have $\bar{C}$ (the Chern number of their groundstate) edge states. Translating everything back to the language of electrons and using the fact that $\bar{j}_{\mu}=\sum_{\alpha}j_{\alpha,\mu}=m j_{e,\mu}$, we will obtain the Chern Simons action level $m$ which is the low energy description of the $1/m$ Laughlin state. 

In conclusion, we have studied the low energy theory of the recently reported fractional quantum Hall effect in the absence of external magnetic field. We have shown that lattice effects can lead to a state similar to the FQH states. We also used the hierarchy method and $K$-theory to generalize these states to other filling fractions. Effective charge, many body Chern number, statistics of quasiparticle and the groundstate degeneracy are discussed as well.  

\section{Acknowledgement} I would like to thank M. Barkeshli, M.S. Vaezi, and R. Asgari for useful discussions.

\end{document}